\documentclass[10pt, a4paper]{article}

\usepackage[final]{lrec-coling2024} 
\usepackage{subcaption}
\usepackage{mathabx}
\usepackage{stfloats}

\title{Evaluating Text-to-Speech Synthesis from a Large Discrete Token-based Speech Language Model}

\name{Siyang Wang, Éva Székely}

\address{
  Division of Speech, Music and Hearing, KTH Royal Institute of Technology, Stockholm, Sweden\\
  \{siyangw,szekely\}@kth.se}

\abstract{
Recent advances in generative language modeling applied to discrete speech tokens presented a new avenue for text-to-speech (TTS) synthesis. These speech language models (SLMs), similarly to their textual counterparts, are scalable, probabilistic, and context-aware. While they can produce diverse and natural outputs, they sometimes face issues such as unintelligibility and the inclusion of non-speech noises or hallucination. As the adoption of this innovative paradigm in speech synthesis increases, there is a clear need for an in-depth evaluation of its capabilities and limitations. In this paper, we evaluate TTS from a discrete token-based SLM, through both automatic metrics and listening tests. We examine five key dimensions: speaking style, intelligibility, speaker consistency, prosodic variation, spontaneous behaviour. Our results highlight the model's strength in generating varied prosody and spontaneous outputs. It is also rated higher in naturalness and context appropriateness in listening tests compared to a conventional TTS. However, the model's performance in intelligibility and speaker consistency lags behind traditional TTS. Additionally, we show that increasing the scale of SLMs offers a modest boost in robustness. Our findings aim to serve as a benchmark for future advancements in generative SLMs for speech synthesis.
 \\ \newline  }

\begin{document}

\maketitleabstract

\section{Introduction}
Generative speech language modeling through next-token prediction on discrete speech tokens \cite{lakhotia2021generative}, referred to as speech language models (SLM) \footnote{Some literature also refers to continuous speech representation models such as WavLM \cite{chen2022wavlm} speech language models. In this paper, we use the term SLM to refer to discrete token-based generative speech language models exclusively.}, was initially proposed as a general speech processing paradigm without the need for text transcription. This was soon adapted to achieve text-to-speech synthesis \cite{wang2023neural, kharitonov2023speak, borsos2023soundstorm, wang2023viola, maiti2023voxtlm, lajszczak2024base, betker2023better}. This paradigm is highly scalable, similar to text-based language models. Several proposed SLMs are trained on speech datasets significantly larger than conventional TTS corpora. It has been shown that SLMs output natural and varied speech samples \cite{betker2023better, kharitonov2023speak}. Recent research suggests that with sufficient training data and model parameter scaling, SLM-based TTS can exhibit what is termed an ``emergent ability" to render appropriate prosody for complex text inputs \cite{lajszczak2024base}. Furthermore, SLMs have showcased in-context learning capability similar to their textual counterparts. In particular, several SLMs boast so-called ``zero-shot TTS" ability, whereby only a few seconds of a speech clip is enough to condition the model to mimic the speaker unseen during training with a reported high level of similarity \cite{wang2023neural,kharitonov2023speak,lajszczak2024base}. SLMs have also demonstrated strong multi-tasking capabilities to solve synthesis-adjacent tasks, such as speech editing \cite{wang2023speechx}, speech-to-speech translation \cite{rubenstein2023audiopalm, wang2023viola}, combining synthesis and recognition into a single model \cite{maiti2023voxtlm}.

However, several important aspects of SLM-based TTS are not thoroughly evaluated in the literature. Questions arise, such as: How meaningful is the variation of the generated output? How do these models handle text inputs with different speaking styles? How do they measure up against traditional TTS systems? Recognizing the growing prominence of this methodology in speech synthesis, we believe that a comprehensive evaluation of a current model is essential, with the objective of shedding light on both its strengths and limitations. This study evaluates five key dimensions: speaking style,
intelligibility, speaker consistency, prosodic variation and spontaneous behaviour. We also publicly  release evaluated audio samples and evaluation code here\footnote{Paper resource page:  \url{https://swatsw.github.io/lrec24_eval_slm/}}.

Our findings indicate that the evaluated SLM generates highly diverse and natural output in terms of prosody and spontaneous behavior. It performs well in both read-speech and conversational speaking styles. However, we find that the model's primary limitations lie in its robustness, as evident in its low intelligibility and speaker consistency.

\begin{figure*}[b]
    \centering
    \includegraphics[width=\textwidth]{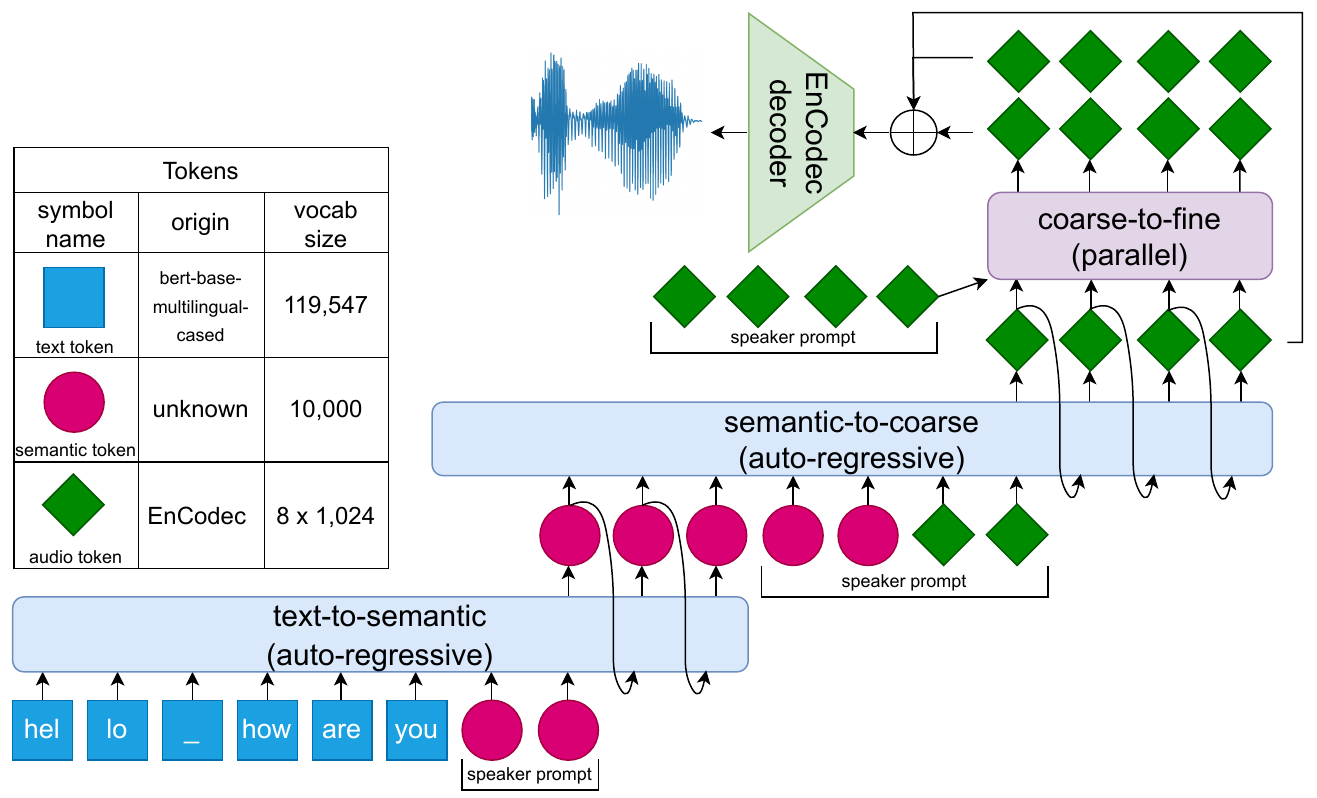}
    \caption{Illustration of Bark. 
    }
    \vspace{-1.5em}
    \label{fig:bark}
\end{figure*}
\vspace{-0.5em}

\section{Evaluated Model: Bark}
Several SLMs capable of TTS have been proposed. Our work evaluates Bark \cite{bark} \footnote{Official repository: https://github.com/suno-ai/bark. Repository commit used in this paper: 599fed0. Model weights retrieved from https://huggingface.co/suno/bark, version number: a3f055a.}. 
There are three reasons for making this choice. First and foremost, both code and weights of Bark are open-sourced. The alternatives are, to the best of our knowledge, not publicly available through code repository or inference API. Secondly, Bark is representative of the current state-of-the-art discrete-token SLM approach to synthesis in terms of model/data scale, model architecture, and training methodology \cite{kharitonov2023speak, borsos2023soundstorm,lajszczak2024base}. This increases the likelihood that our findings may apply to other models not specifically evaluated in this study. Lastly, we deduce from its output that Bark is trained on a mixed-style dataset, while prior SLMs are trained on either read/audiobook-only corpora \cite{kharitonov2023speak}, or spontaneous/conversational-only corpora \cite{borsos2023soundstorm,chen2023vector}. This allows us to assess Bark’s performance across different speaking styles and investigate the impact of employing a mixed-style dataset for SLM training.

Even though Bark is open-sourced, no official technical document was released. To bridge this gap of knowledge, we derive the details of the model presented in this section directly from the model's official codebase. Should any discrepancies arise between our description and the actual implementation, the latter should be considered authoritative.

\subsection{Model Architecture}
\label{sec:model_arch}
Bark consists of 3 levels of discrete-token speech language model as illustrated in Figure \ref{fig:bark}. The three levels operate in succession. 

The first level, called \textit{text-to-semantic}, is a decoder-only auto-regressive transformer that takes in text tokens, in this case text language model tokens, additionally encodes a given speaker prompt in the form of semantic tokens as well as current generated sequence of semantic tokens, and outputs next semantic-token distribution. The semantic token vocabulary has a size of 10,000 and is believed to be similar to semantic tokens used in AudioLM \footnote{Suno-AI acknowledged AudioLM as reference for building Bark \cite{bark}.} \cite{borsos2023audiolm} and SPEAR-TTS \cite{kharitonov2023speak}. Semantic tokens serve as a middle representation between text and more granular audio tokens.

The second level, \textit{semantic-to-coarse}, is another decoder-only auto-regressive transformer. It takes generated semantic token sequence from the first level, additionally encodes a speaker prompt in the form of combined semantic and audio tokens along with current generated sequence of audio tokens, and outputs distribution for the next audio token. The audio tokens are from an audio compression model EnCodec \cite{defossez2023high}, which compresses speech audio into 8 separate code books each of size 1024 at 75Hz. Semantic-to-coarse transformer generates tokens for the first 2 code books or the \textit{coarse tokens}. 

The third level, \textit{coarse-to-fine}, takes in generated coarse tokens from the second level, and generates tokens for the other 6 code books of the EnCodec model or the \textit{fine tokens}. The generation is conditioned on a speaker prompt in the form of fine tokens. This level is different from the first two in that it is an encoder-only transformer, i.e. it generates in parallel and not auto-regressively. After running the third level, the resulting sequence for all 8 code books of the EnCodec model is fed into the corresponding EnCodec decoder to obtain the final waveform. 

In order to generate speech from a given speaker, Bark is prompted at all three levels with tokens from the given speaker. This is a standard technique in SLMs \cite{wang2023neural,kharitonov2023speak}, often referred to as ``zero-shot" synthesis since only a short clip from a speaker is needed to condition the synthesis on that speaker. However, it is not clear if the prompt tokens provide additional context besides speaker identity. For example, if the prompt tokens are from the prior utterance, does the model take into account the prosody or semantics in the prompt to generate more meaningful speech? We test this hypothesis by replacing the speaker prompt at semantic-to-coarse level with encoded semantic tokens from prior utterance. This model condition is tested in the listening tests in Section \ref{sec:eval_listen}.

\subsection{Baseline: VITS}
We use a multi-speaker VITS \cite{kim2021conditional} \footnote{Implementation: https://github.com/coqui-ai/TTS} trained on VCTK dataset \citetlanguageresource{veaux2017cstr} as baseline. Our justification for this choice is two-fold. Firstly, this model is representative of current state-of-the-art TTS models in several regards: architecture, input representation (phonemes), and training corpus. The model is designed to be probabilistic and is therefore able to address the one-to-many problem in TTS, which is an important aspect of this evaluation. Secondly, VITS models phoneme duration explicitly, achieving a level of intelligibility that can be seen as a top line for TTS models. Thus, we can effectively assess Bark's intelligibility through comparison with VITS. We note that VITS differs from Bark in several dimensions, including model architecture, data representation, and training data. Nonetheless, we designed extensive evaluations that are fair from the black-box usability point of view.

\begin{table}
\centering
\begin{tabular}{llr}
\hline
\textbf{Model} & \textbf{Param.} & \textbf{RTF}\\
\hline
\hline
Bark-small:                 & 389M$\quad$ & 0.660$\pm$0.01\\
$\drsh$ text-to-semantic    & $\drsh$193M & 0.171$\pm$0.00\\
$\drsh$ semantic-to-coarse  & $\drsh$104M &  0.479$\pm$0.00\\
$\drsh$ coarse-to-fine      & $\drsh$\;\;92M &0.009$\pm$0.00\\
\hline
Bark-base:                  & 1B$\quad$ & 1.250$\pm$0.01\\
$\drsh$ text-to-semantic    & $\drsh$446M & 0.331$\pm$0.00\\
$\drsh$ semantic-to-coarse  & $\drsh$328M & 0.911$\pm$0.00\\
$\drsh$ coarse-to-fine      & $\drsh$312M  & 0.027$\pm$0.01\\
\hline
VITS                        & 40M & 0.049$\pm$0.01\\
\hline
\hline
\end{tabular}
\caption{Parameter count and real-time-factor.}
\label{tab:arch_param_runtime}
\vspace{-1.0em}
\end{table}

\subsection{Parameter Count and Runtime}
To clearly see the scale comparison between the models, we present model size and synthesis real-time-factor (RTF) in Table \ref{tab:arch_param_runtime}. The Bark models have a much bigger parameter count than VITS. Even the smaller Bark-small has 10 times the parameter count of VITS. This is partially due to the ultra-large vocabulary size used by Bark as it encodes text using a variation of word-piece used in large text language models, while a conventional phoneme-based TTS like VITS only has a small phoneme vocabulary. On the runtime side, Bark models are much slower to run compared to VITS. The main contributing factor is Bark's autoregressive nature. VITS generates in parallel by a fully-convolutional network. Furthermore, the much higher parameter count in Bark could also have contributed to slower synthesis speed.

\section{Evaluation Method}
In this section, we present our methodology for evaluating Bark on the following dimensions: speaking style, intelligibility, speaker consistency, prosody variation, spontaneous behaviour, subjective listening impression. We elaborate on the rationale and specific methods employed for each of these dimensions. Speaking style is interwoven with all other evaluated dimensions as we utilize two types of text inputs, corresponding to two speaking styles: reading and conversational, in all evaluations.

In all evaluations besides the listening tests, we use the set of 10 speaker prompts corresponding to 10 different speakers provided in the Bark repository to guide model synthesis, which we refer to as Bark speaker 0 to 9. To reduce the potential impact of speaker differences on our evaluation, we match each Bark speaker with a similar speaker from the VITS model’s VCTK training dataset. This matching is based on the speaker similarity metric, as detailed in the speaker consistency evaluation in section \ref{sec:eval_speaker_consistency}. This results in 10 Bark speakers and their corresponding 10 VITS speakers for all evaluations, except for listening tests for which the speaker selection process is described in section \ref{sec:eval_listen:speaker_selection}.

\subsection{Text Input: Two Speaking Styles}
To investigate model performance across different speaking styles, we utilize two distinct sets of text inputs corresponding to two speaking styles: read-speech text from LibriTTS \citetlanguageresource{zen2019libritts} and conversational text from DailyDialog \citetlanguageresource{li2017dailydialog}. 
\label{sec:corpora}
\subsubsection{Read-Speech Text: LibriTTS}
To evaluate read speech synthesis, we use text from the LibriTTS \citetlanguageresource{zen2019libritts} corpus as input text prompts to the TTS models. This corpus consists of open-access books read aloud by multiple speakers. Utterance ordering in the original book is carefully recorded in the metadata, thus it is possible to retrieve the preceding utterance. We utilize this aspect of the corpus to evaluate synthesized speech in the context of its preceding utterance. We randomly sampled 2,400 utterance texts from the official test split for all experiments.

\subsubsection{Conversational Text: DailyDialog}
To evaluate conversational synthesis, we use the DailyDialog \citetlanguageresource{li2017dailydialog} corpus as text input. It consists of written single-themed conversations between two interlocutors without clearly defined roles. We use 2,400 randomly chosen utterances from the official validation split of the corpus for all experiments.

\subsection{Intelligibility}
\label{sec:eval_intelligibility}
We apply Automatic Speech Recognition (ASR) to synthesized samples. The word error rate (WER) between ASR output and TTS input is used as a proxy for intelligibility of synthesized speech as prior study has found that ASR accuracy is correlated with human transcription accuracy \cite{taylor2021confidence} in evaluating TTS. The ASR used is Whisper \cite{radford2023robust} base model specifically trained for English.

\subsection{Speaker Consistency}
\label{sec:eval_speaker_consistency}
We subjectively found that the Bark output has minor speaker drift such that several sampled outputs from the same speaker sound like different speakers. To probe how consistent is Bark in this regard, we use speaker similarity score from a strong speaker identification model ECAPA-TDNN \cite{desplanques2020ecapa}\footnote{Implementation: https://speechbrain.github.io/}, calculated as cosine-distance of its embedding space, as speaker consistency metric for synthesized speech samples. Taking synthesized samples from the same speaker, we calculate speaker similarity scores for all enumerated sample pairs, we then calculate mean and standard deviation of the scores. These two statistics are used as the metrics for speaker consistency within the same speaker. A high mean and low standard deviation indicates high speaker consistency, and vice versa for low speaker consistency. We also calculate inter-speaker similarity as we hypothesize that as the model drifts away from the given speaker it drifts towards another speaker in its distribution. Thus, we expect to see that the model with low within-speaker speaker consistency to have higher inter-speaker similarity. 

\subsection{Prosodic Variation}
\label{sec:eval_prosody}
One of the predominant challenges in text-to-speech is known as the 'one-to-many problem'. The same input text often has multiple reasonable realizations in the speech domain. Consequently, a good TTS model should be able generate range of plausible prosodic interpretations given the same input text. We will assess the how plausibly the prosody is in \ref{sec:eval_listen} Listening Test. This part of the evaluation is focused on how varied is the prosody when the system is presented with identical sentence inputs. Here we use a quantitative evaluation method, based on automatically extracted prosodic features. The suitability of the generated prosody will be evaluated in the listening test. The prosody dimensions measured are fundamental frequency f0 and speech rate. f0 is measured using YIN algorithm \cite{de2002yin} \footnote{Implementation: https://github.com/brentspell/torch-yin/tree/main}. Speech rate is measured as syllables per second. We found that ASR transcription is more accurate in counting syllables than text input, since there could be significant discrepancy between synthesized speech and input text especially in Bark models. We thus opted to use ASR transcription to count syllables when measuring speech rate.

After measuring the two prosody metrics on each speech sample, we calculate their standard deviations by utterance text. Each utterance text is synthesized with different speakers and has several sampled synthesis for the same speaker. We aggregate all speakers and all sampled synthesis per speaker for the same utterance test into a list, and calculated the standard deviation of this list. We would then summarize standard deviations for all utterances by model to understand between-model difference in the amount of prosody variation. Through informal subjective listening tests, we found that Bark samples are more varied for the same text input. We therefore hypothesize that it would have high standard deviation in both prosody measures for most samples.

\subsection{Spontaneous Behaviour}
\label{sec:eval_spont}
We measure two spontaneous behaviours present in the synthesis output:  the insertion of fillers and pauses. Both can be measured automatically. We count the number of fillers through ASR output. Here we define fillers as speech tokens that do not add to the propositional content of the message and use the following set: ``um", ``uh", ``ah", ``mm", ``hm", ``hmm", ``huh", ``er", ``eh", ``mhm", ``mmh". The first two are most common. We tuned the optional prompt input in Whisper to elicit better filler transcription on a small test set. It still misses some filler occurrences, so this count can be seen as under-measured but the under-measurement should equally affect all evaluated models. 

The pauses are measured as non-speech segments within the speech utterance. We first run the speech samples through a Voice Activity Detector \footnote{https://github.com/snakers4/silero-vad} to get speech segments. The spaces between them are non-speech segments and are treated as pauses, the total length of which is the pause length of the speech segment. 

\subsection{Listening Tests}
\label{sec:eval_listen}
We conduct listening tests to assess two aspects of synthesized speech with human listeners: overall naturalness and contextual suitability. For each test we conduct two separate parts in the two speaking styles. We assess 3 samples of each input sentence from each evaluated model. All tests follow the MOS listening test specification in ITU standard P.800 \cite{itu1996telephone}. The main deviation from the standard is the instruction texts given to listeners, which are specific to the two tests. We use only Bark-base in these tests as comparing scales is not the main focus of these tests.

\subsubsection{Test 1: MOS-N(aturalness)}
The MOS-N test is similar to how standard speech naturalness assessment is done in TTS model development. We ask the listeners to rate "How natural does the speech sample sound?". It is a single combined measure of different aspects of the speech audio, including signal quality, noise level, and prosody naturalness. We use this test to benchmark Bark's synthesis output in the same way as standard TTS research practice.

\subsubsection{Test 2: MOS-Contexual-S(uitability)}
As mentioned in Section \ref{sec:eval_prosody}, we also want to assess how Bark's high variation in prosody and spontaneous behavior are perceived by the human listener. We hypothesize that such high variation makes Bark's synthesis output more suitable for a variety of specific contexts. This contrasts with conventional TTS systems which may come across as monotonous and less tailored for distinct scenarios due to their limited prosodic variance. Consider a brief dialogue: "A: How are you? B: I'm fine.". The response "I'm fine." can be articulated in myriad ways to imply different sentiments. Each rendition might emphasize unique prosodic elements or spontaneous nuances to relay the intended meaning appropriate to the context. Prior studies have validated that such contextual suitability can be abundantly detected in listening tests similar to ours \cite{wallbridge2021s}.  We hypothesize that Bark can synthesize such variations, which would often align with human expectations in terms of contextual appropriateness or suitability. Conversely, VITS, with its constrained prosodic range, might only match a singular context or occasionally might not align with any. 

Similarly to the MOS-N test, we conduct separate parts of MOS-Context-S test in the two speaking styles: read-speech and conversational. For the read-speech evaluation, we leverage the LibriTTS corpus \citetlanguageresource{zen2019libritts}, where all sentences orginate from books. Consequently, the immediate preceding sentence serves as context to a speech sample. In the conversational test, we use the preceding dialogic exchange (i.e. the turn of the interlocutor) in DailyDialog \citetlanguageresource{li2017dailydialog} as context. Both the context and the input are shown to the listeners. We ask the listeners to rate "How suitable does the speech sample sound in the context?".

\subsubsection{Speaker Selection for Listening Tests}
\label{sec:eval_listen:speaker_selection}
We selected two Bark speakers for listening tests: speaker 6, which exhibits a spontaneous conversational style with a male voice; and speaker 9 which is less spontaneous as shown in Table \ref{tab:spontaneous}, but has varied prosody, and sounds like a female voice. For each of the two speaking styles, we randomly sampled a VITS speaker as baseline comparison. As mentioned at the end of Section \ref{sec:model_arch}, we test an additional model variation of Bark where the speaker prompt is replaced by a context prompt from the prior utterance. This variation is applied to Bark-9 for LibriTTS and Bark-6 for DailyDialog.  Thus, we have the following set of model-speakers for LibriTTS: Bark-6, Bark-9, Bark-context-9 and VITS-p243; and the following set for DailyDialog: Bark-6, Bark-9, Bark-context-6, and VITS-p247.

In order to neutralize any perceptual biases stemming from voice timbre variations across the speakers, we apply voice conversion to all test speech samples using a third VITS speaker. This speaker is chosen based on equidistant speaker embedding metric\footnote{Calculated in the same way as Section \ref{sec:eval_speaker_consistency}.} to the three evaluated speakers, ensuring a roughly equivalent challenge in converting each of the three primary speakers to the target VITS speaker. We use a state-of-the-art voice conversion model FreeVC \cite{li2023freevc} \footnote{Implementation: https://github.com/coqui-ai/TTS}.

\section{Results}
\begin{figure*}[b]
    \includegraphics[width=\textwidth]{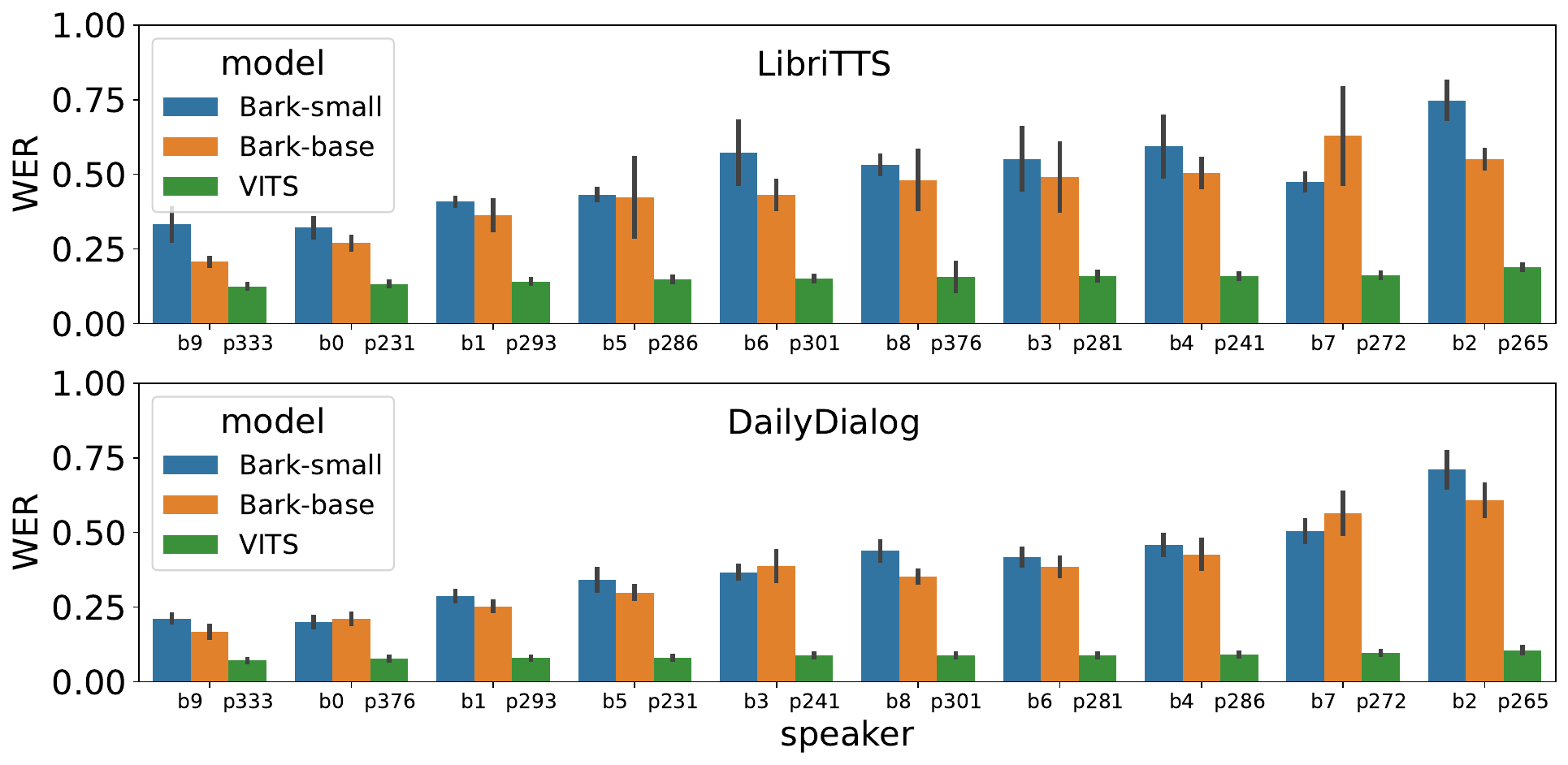}
    \caption{Word error rate (WER). Speaker denotation is b[x] for Bark speakers, p[x] for VITS speakers. Speaker order is based on ascending-sorted WER.}
    \label{fig:wer}
\end{figure*}

\subsection{Intelligibility}
For automatically assessing intelligibility, we employ the ASR word error rate (WER) from Whisper as an indicative measure, as detailed in Section \ref{sec:eval_intelligibility}. Both the ground-truth TTS input and the Whisper transcriptions are normalized by removing punctuations and converting to lowercase. The results are shown in Figure \ref{fig:wer}. 

Two prominent observations emerge from this analysis: 1) Bark consistently registers a higher WER in comparison to VITS across both corpora. The Bark speaker with the lowest WER still surpasses the highest WER recorded by a VITS speaker. 2) When comparing two Bark models at different scales the larger model consistently outperforms its smaller counterpart in terms of average intelligibility. This observation holds true for 9 out of 10 speakers in LibriTTS and 7 out of 10 in DailyDialog. 

\begin{figure*}
    \centering
    \subcaptionbox{Bark-small\label{fig:speaker_bark_small}}{
      \includegraphics[width=.33\linewidth]{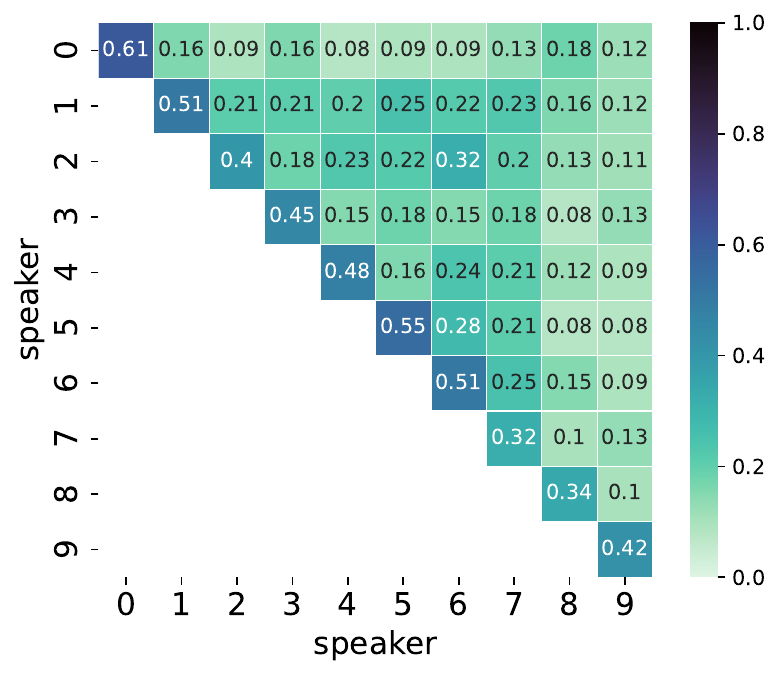}}%
    \subcaptionbox{Bark-base\label{fig:speaker_bark_base}}{
      \includegraphics[width=.33\linewidth]{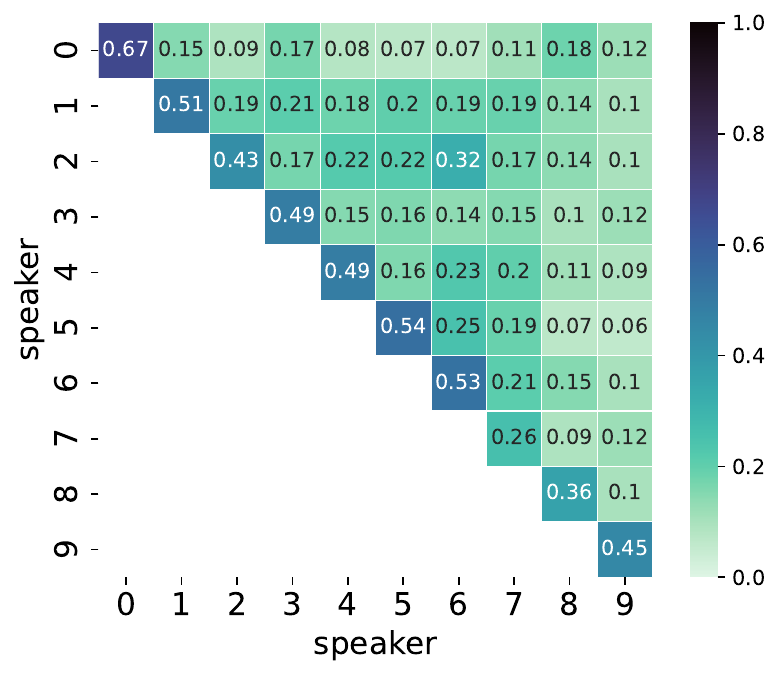}}%
    \subcaptionbox{VITS\label{fig:speaker_vits}}{
      \includegraphics[width=.33\linewidth]{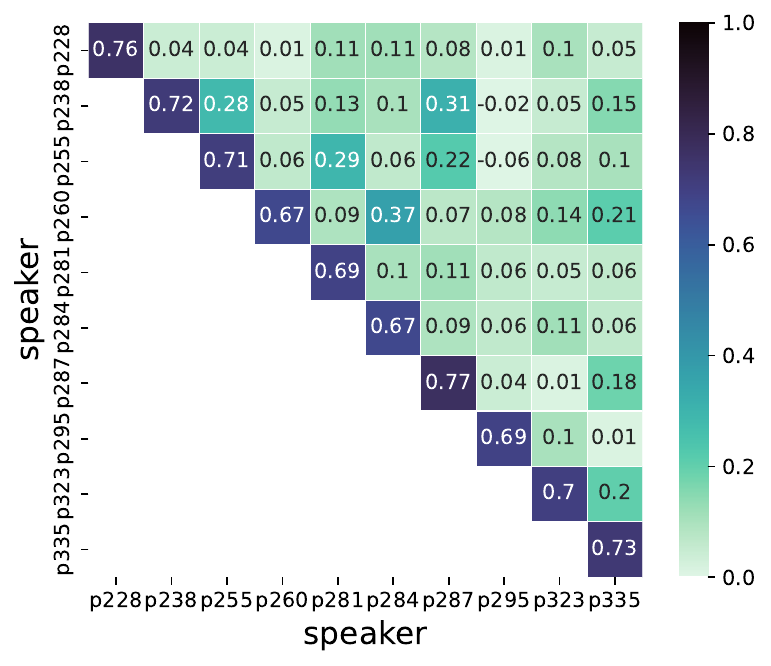}}
    \caption{Speaker similarity metrics. VITS speakers are randomly selected to match the number of speakers in Bark for clear comparison. Color corresponds to the mean similarity calculated by a speaker embedding model ECAPA-TDNN \cite{desplanques2020ecapa}.
    }
    \label{fig:speaker}
\end{figure*}

\subsection{Speaker Consistency}
Speaker similarity matrices are shown in Figure \ref{fig:speaker}. Ideally, the diagonal cells, which denote intra-speaker similarity, should exhibit high values. This would indicate that when models are conditioned on a specific speaker, the resulting samples consistently align with that speaker's profile. While it's evident that diagonal cells generally manifest heightened speaker similarity compared to the off-diagonal cells (representing inter-speaker similarity), this trend is much more pronounced in VITS.

For several Bark speakers, there's a concerning observation: the off-diagonal cells showcase intensities nearly matching the diagonal cells, especially in speaker 2 and 7. This implies that the samples synthesized from a specific Bark speaker bear a similarity to samples from other speakers almost as much as they do to their own. Simply put, Bark appears to struggle in maintaining robust speaker conditioning. The synthesized output might not always faithfully replicate the conditioned speaker's attributes. Furthermore, when Bark diverges from the conditioned speaker, it tends to gravitate towards other speakers within its training set. This is evidenced by the heightened similarity values in the off-diagonal cells, which correspond to other modeled speakers. Bark-base has slightly more robust intra-speaker consistency as evident in higher values in diagonal cells in 7 out of 10 speakers, otherwise the two sizes of Bark models bear little difference.

\begin{figure*}
    \centering
    \includegraphics[width=1.0\linewidth]{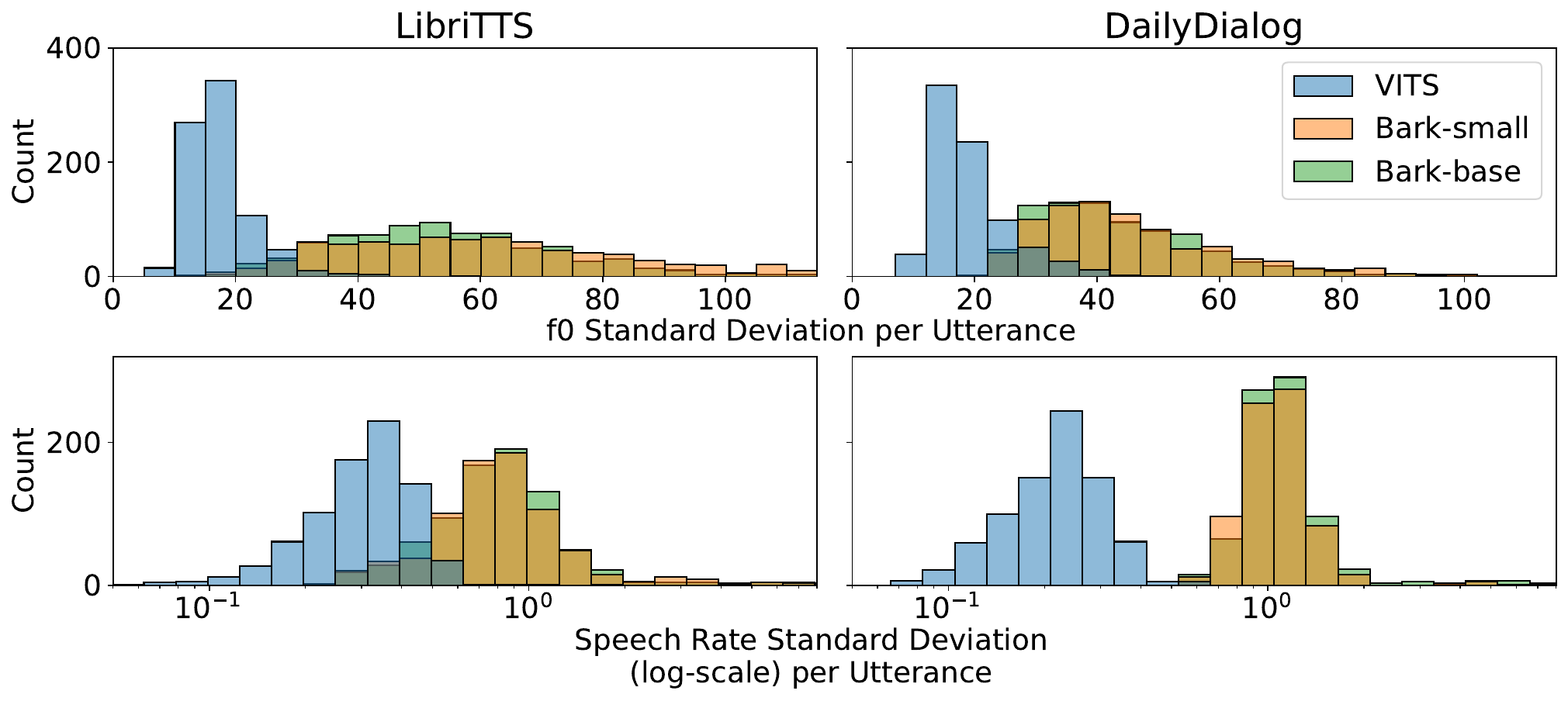}
        \vspace{-2.0em}
    \caption{Prosodic variation.}
    \label{fig:prosody}
\end{figure*}

\subsection{Prosodic Variation}
\label{sec:results_prosody}
As detailed in Section \ref{sec:eval_prosody}, we measured two prosodic dimensions, f0 and speech rate. We then calculated standard deviation of the measured values by utterance. The results are visualized in Figure \ref{fig:prosody}. Our hypothesis is that Bark generates more varied prosody than VITS. This is indeed the case as evident in Bark's higher standard deviation in both measured prosody metrics compared to VITS. The results also show that the spread of standard deviation values is more expansive for Bark. This indicates not only a generally more varied prosody, but also a more fluctuating degree of prosodic variation across different utterances when compared to VITS. The distributions of small and base Bark models are similar suggesting that scale does not affect level of prosodic variation in SLM.

\begin{table*}
\centering
\begin{tabular}{l|llllllllll}
\hline
\bf{Bark speakers} &\bf{0}&\bf{1}&\bf{2}&\bf{3}&\bf{4}&\bf{5}&\bf{6}&\bf{7}&\bf{8}&\bf{9}\\
\hline
\bf{filler count} &0.0&0.0&0.06&0.04&0.02&0.0&\bf{0.55}&0.06&0.01&0.0 \\
\bf{pause duration}&0.29&0.27&0.35&\bf{0.65}&0.37&0.38&0.61&0.38&0.3&0.46 \\
\hline
\hline
\bf{VITS speakers} &\bf{p286}&\bf{p301}&\bf{p231}&\bf{p333}&\bf{p272}&\bf{p241}&\bf{p376}&\bf{p281}&\bf{p293}&\bf{p265} \\
\hline
\bf{filler count} &0.0&0.0&0.0&0.0&\bf{0.01}&0.0&0.0&0.0&0.0&0.0 \\
\bf{pause duration}&0.18&0.19&0.21&\bf{0.22}&0.2&0.21&0.18&0.19&0.21&0.19 \\
\hline
\end{tabular}
\caption{Spontaneous behaviour metrics. Mean across DailyDialog utterances is shown.}
\label{tab:spontaneous}
\end{table*}

\subsection{Spontaneous Behaviour}
We measured spontaneous behaviours as detailed in Section \ref{sec:eval_spont}, insertion of fillers and pauses. Results in DailyDialog are shown in Table \ref{tab:spontaneous}, results are similar in LibriTTS. For Bark speakers, we see that speaker 6 inserts most fillers per utterance at around 0.55 on average, while other speakers like 0 and 9 do not insert fillers at all. None of the VITS speakers insert fillers as expected, since the model's explicit duration modeling strongly discourages it from deviating from the input text. Additionally, Bark tends to insert longer pauses on average compared to VITS. Both pause and filler insertion among Bark speakers also display considerable variance, with standard deviation as high as 2.3 for filler count and 1.1 for pause duration in speaker 6 and 3, while VITS only reaches 0.1 and 0.5 in corresponding measures. Bark results in Table \ref{tab:spontaneous} are from Bark-base. Bark-small obtains similar results.

\begin{table*}
\centering
\vspace{-1.0em}
\begin{tabular}{@{}l|c|c|c|c@{}}
 & \multicolumn{4}{c}{\textbf{LibriTTS}} \tabularnewline
 & \multicolumn{2}{c}{MOS-Naturalness} & \multicolumn{2}{c}{MOS-Context-Suitability}
 \tabularnewline
& Mean & Range & Mean & Range
  \tabularnewline
\hline
 VITS-p243  & 2.86 $\pm$ 0.18 & 1.10 $\pm$ 0.17 & 2.75 $\pm$ 0.17 & 1.06 $\pm$ 0.21 \tabularnewline
 Bark-6     & 3.34 $\pm$ 0.14 & 1.09 $\pm$ 0.19 & 2.64 $\pm$ 0.18 & 1.24 $\pm$ 0.20 \tabularnewline
 Bark-9     & 3.36 $\pm$ 0.13 & 0.95 $\pm$ 0.21 & 3.49 $\pm$ 0.17 & 1.24 $\pm$ 0.22 \tabularnewline
 Bark-context-9 & \bf{3.56 $\pm$ 0.16} & \bf{1.19 $\pm$ 0.19} & \bf{3.58 $\pm$ 0.19} & \bf{1.27 $\pm$ 0.28} \tabularnewline
\hline
  & \multicolumn{4}{c}{\textbf{DailyDialog}} \tabularnewline
 & \multicolumn{2}{c}{MOS-Naturalness} & \multicolumn{2}{c}{MOS-Context-Suitability}
 \tabularnewline
 & Mean & Range & Mean & Range
  \tabularnewline
\hline
 VITS-p247  & 3.39 $\pm$ 0.14 & 0.98 $\pm$ 0.17 & 3.33 $\pm$ 0.14 & 1.04 $\pm$ 0.17 \tabularnewline
 Bark-6   & 3.40 $\pm$ 0.12 & \bf{1.32 $\pm$ 0.20} & 3.34 $\pm$ 0.13 & \bf{1.29 $\pm$ 0.22} \tabularnewline
 Bark-9   & \bf{3.53 $\pm$ 0.13} & 1.21 $\pm$ 0.18 & \bf{3.48 $\pm$ 0.18} & 1.28 $\pm$ 0.22 \tabularnewline
 Bark-context-6 & 3.39 $\pm$ 0.13 & 1.22 $\pm$ 0.17 & 3.25 $\pm$ 0.11 & 1.25 $\pm$ 0.15 \tabularnewline

\end{tabular}
\caption{Listening test results. 5-scale Mean-Opinion-Score (MOS). Each utterance has 3 samples. ``Mean" corresponds to the mean of the 3 samples, while ``Range" corresponds to (max-min) of the 3. These two metrics are then calculated mean and confidence interval (p=0.05) across utterances. }
\label{tab:mos}
\vspace{-1.5em}
\end{table*}

\subsection{Listening Tests}
We conducted listening tests detailed in Section \ref{sec:eval_listen}. Initial observations from a pilot study revealed that Bark output with high WER are easy to detect by listeners and are rated the lowest, mostly because they contain clearly audible synthesis mistakes like non-speech noise or strange voicing. Our intelligibility results in Section \ref{sec:eval_intelligibility} already revealed that Bark is not robust in this regard. Therefore, including samples with compromised intelligibility in the listening test could introduce a confounding variable, potentially obfuscating genuine insights about the TTS. To mitigate this, we excluded samples with a WER exceeding 0.1, applying this criterion uniformly to both Bark and VITS outputs. After this filtering, we retained 33 utterances for LibriTTS and 39 from DailyDialog \footnote{Audio samples:  \url{https://swatsw.github.io/lrec24_eval_slm/}}. Each utterance has 3 intelligible samples from each model.

We recruited 30 native English speakers for each of the 4 tests (2 parts in both MOS-N and MOS-Context-S) on crowd-sourcing platform Prolific \footnote{prolific.com}. Each participant was tasked with rating between 50 to 60 samples, ensuring a balanced representation across utterances and model-speakers. Each sample received 5 ratings on average, the mean of the ratings is used as the MOS for that sample. For each utterance, we calculate two metrics, the mean MOS of the 3 samples, and the range (max - min) of the 3. The results are shown in Table \ref{tab:mos}.

In mean MOS, we find Bark speaker 9 to be strong in all test scenarios while VITS speakers are rated the lowest or the second lowest. This suggests that, in both read-speech and conversational styles, Bark synthesis is more natural and more contextually appropriate to VITS. Bark-context, the Bark variation with prior utterance as prompt, has the highest rating in LibriTTS but not in DailyDialog. Therefore it is not clear whether providing prior utterance as context through prompt tokens improves synthesis. This puts into question if speech language model TTS incorporates prosodic or semantics context in the prompt. We apply Tukey HSD test (with FWER=0.05) on all pair-wise model-speaker comparisons in all 4 tests and find that Bark-context-9 is significantly better than VITS-p243 in both MOS-Naturalness and MOS-Context-Suitability in LIibriTTS, however no significance is found in DailyDialog. This suggests that Bark's advantage over VITS is stronger in read-speech synthesis than in conversational synthesis.

The MOS range reveals how varied the ratings are for the 3 samples of each utterance. As results in prosodic variation (Section \ref{sec:results_prosody}) have shown, Bark produces more varied samples with the same input text compared to VITS. This is again evident in the MOS range statistics where Bark models have clearly higher MOS range across the test sets. We further validate this by comparing MOS variances with Fligner-Killeen test which shows that the Bark speaker with the highest variance in each test is significantly more varied than VITS. 


\section{Discussion}
Admittedly, our evaluation was centered on a single SLM, primarily due to the limited public access of similar models. This focus may limit the generalizability of our findings across other SLMs. When comparing our results with existing literature, it becomes apparent that there are significant differences in performance between models. For instance, in terms of intelligibility, BASE-TTS \cite{lajszczak2024base} reported a 6.5 WER, while Bark obtained 19.2 on the same set of text inputs. Similarly, VoxtLM \cite{maiti2023voxtlm} reported a modest 8.8 Character Error Rate (CER). Thus, both BASE-TTS and VoxtLM are significantly more robust SLMs in terms of TTS intelligibility, highlighting substantial variability between SLM models. Another example of such between-model difference is how model performance changes with increased scale. While we found that larger Bark models are more robust in terms of intelligibility and speaker consistency, VoxtLM reported that increased model size did not result in increased synthesis robustness, as their 1.3B model obtained worse CER than their 350M model. They found that instead the speech token vocabulary size is more important for high TTS intelligibility as increased token vocabulary increased TTS intelligibility.

Nonetheless, our findings resonate with patterns observed in the current literature. For instance,  BASE-TTS \cite{lajszczak2024base} demonstrated that listener preferences tend to favor larger models trained on more data. We observed a similar trend in Bark, where both intelligibility and speaker consistency improved as the model scale increased. We hope that future research will leverage our evaluation framework to explore additional emerging SLMs to gain a deeper understanding of the dynamics between model scale, training data volume, and overall TTS performance in SLMs.



A distinguishing characteristic of Bark is its capability for multi-lingual synthesis. Besides English, Bark provides a range of other languages. Enabling multi-lingual synthesis in a single model is a powerful feature not typically seen in traditional TTS models. This achievement has also been mirrored by several other SLMs \cite{wang2023viola,rubenstein2023audiopalm}.
Given the observed speaker inconsistencies within Bark, it is plausible that languages might influence one another within the model, leading to intriguing interactions that warrant investigation in the future.

\section{Conclusions}
We evaluated TTS from a large discrete token-based speech language model (SLM), Bark, along the dimensions of speaking style, intelligibility, speaker consistency, prosody variation, spontaneous behaviour. Through a series of carefully designed automatic quantitative measurements and subjective listening tests, we find that Bark generates highly variable and natural prosody as well as spontaneous behaviours. However, it falls short in robustness when compared with a conventional TTS model, particularly in terms of intelligibility and speaker consistency. Interestingly, we observed that augmenting the model's scale marginally enhances its robustness, suggesting that scaling might be a promising avenue to increase the robustness of SLMs. We believe that our findings can serve as a benchmark for future progress in the development of generative SLMs for synthesis.

\section{Acknowledgements}
This work is supported by Digital Futures project Advanced Adaptive Intelligent Systems (AAIS), and the Swedish Research Council project
Perception of speaker stance (VR-2020-02396).

\nocite{*}
\section{Bibliographical References}\label{sec:reference}

\bibliographystyle{lrec-coling2024-natbib}
\bibliography{my_bib}

\section{Language Resource References}
\label{lr:ref}
\bibliographystylelanguageresource{lrec-coling2024-natbib}
\bibliographylanguageresource{languageresource}

\end{document}